\documentclass[twocolumn]{aastex631}

\usepackage{amsmath}
\usepackage[flushleft]{threeparttable}
\usepackage{soul}
\usepackage{ulem}

\begin{document}

\title{Scintillation velocity and arc observations of FRB~20201124A}
\shorttitle{Scintillation properties of FRB~20201124A}

\correspondingauthor{Ziwei Wu, Weiwei Zhu} 
\email{wuzw@bao.ac.cn, zhuww@nao.cas.cn}

\shortauthors{Wu et al.}

\author[0000-0002-1381-7859]{Ziwei Wu}
\affiliation{National Astronomical Observatories, Chinese Academy of Sciences, Beijing 100101, China}

\author[0000-0001-5105-4058]{Weiwei Zhu}
\affiliation{National Astronomical Observatories, Chinese Academy of Sciences, Beijing 100101, China}
\affiliation{Institute for Frontier in Astronomy and Astrophysics, Beijing Normal University, Beijing 102206, China}

\author[0000-0002-9725-2524]{Bing Zhang}
\affil{Nevada Center for Astrophysics, University of Nevada, Las Vegas, NV 89154, USA}
\affil{Department of Physics and Astronomy, University of Nevada, Las Vegas, NV 89154, USA}

\author[0000-0002-0475-7479]{Yi Feng}
\affiliation{Research Center for Astronomical Computing, Zhejiang Laboratory, Hangzhou 311100, China}

\author[0000-0002-9274-3092]{JinLin Han}
\affiliation{National Astronomical Observatories, Chinese Academy of Sciences, Beijing 100101, China}
\affiliation{School of Astronomy, University of Chinese Academy of Sciences, Beijing, China}

\author[0000-0003-3010-7661]{Di Li}
\affiliation{National Astronomical Observatories, Chinese Academy of Sciences, Beijing 100101, China}
\affiliation{NAOC-UKZN Computational Astrophysics Centre, University of KwaZulu-Natal, Durban 4000, South Africa}
\affiliation{Research Center for Intelligent Computing Platforms, Zhejiang Laboratory, Hangzhou 311100, China}

\author[0000-0001-7931-0607]{Dongzi Li}
\affiliation{Cahill Center for Astronomy and Astrophysics, MC 249-17 California Institute of Technology, Pasadena CA 91125, USA}

\author[0000-0002-4300-121X]{Rui Luo}
\affiliation{Department of Astronomy, School of Physics and Materials Science, Guangzhou University, Guangzhou 510006, China}

\author[0000-0001-6651-7799]{Chenhui Niu}
\affiliation{Institute of Astrophysics, Central China Normal University, Wuhan 430079, China}

\author[0000-0001-8065-4191]{Jiarui Niu}
\affiliation{School of Astronomy, University of Chinese Academy of Sciences, Beijing, China}
\affiliation{National Astronomical Observatories, Chinese Academy of Sciences, Beijing 100101, China}

\author[0000-0002-9434-4773]{Bojun Wang}
\affiliation{National Astronomical Observatories, Chinese Academy of Sciences, Beijing 100101, China}

\author[0000-0003-4157-7714]{Fayin Wang}
\affiliation{School of Astronomy and Space Science, Nanjing University, Nanjing 210093, China}
\affiliation{Key Laboratory of Modern Astronomy and Astrophysics (Nanjing University), Ministry of Education, China}

\author[0000-0002-3386-7159]{Pei Wang}
\affiliation{National Astronomical Observatories, Chinese Academy of Sciences, Beijing 100101, China}

\author[0000-0001-9036-8543]{Weiyang Wang}
\affiliation{School of Astronomy and Space Science, University of Chinese Academy of Sciences, Beijing 100049, China}

\author[0000-0002-5031-8098]{Heng Xu}
\affiliation{National Astronomical Observatories, Chinese Academy of Sciences, Beijing 100101, China}

\author[0000-0001-6374-8313]{Yuanpei Yang}
\affiliation{South-Western Institute For Astronomy Research, Yunnan University, Yunnan 650504, China}

\author[0000-0002-8744-3546]{Yongkun Zhang}
\affiliation{National Astronomical Observatories, Chinese Academy of Sciences, Beijing 100101, China}
\affiliation{School of Astronomy, University of Chinese Academy of Sciences, Beijing, China}

\author[0000-0002-6423-6106]{Dejiang Zhou}
\affiliation{National Astronomical Observatories, Chinese Academy of Sciences, Beijing 100101, China}
\affiliation{School of Astronomy, University of Chinese Academy of Sciences, Beijing, China}

\author[0009-0009-8320-1484]{Yuhao Zhu}
\affiliation{National Astronomical Observatories, Chinese Academy of Sciences, Beijing 100101, China}
\affiliation{School of Astronomy, University of Chinese Academy of Sciences, Beijing, China}

\author[0000-0003-0471-365X]{Can-Min Deng}
\affiliation{Guangxi Key Laboratory for Relativistic Astrophysics, Department of Physics, Guangxi University, Nanning 530004, China}

\author[0000-0001-5662-6254]{Yonghua Xu}
\affiliation{Yunnan Observatories, Chinese Academy of Sciences, Kunming 650011, China}

\collaboration{30}{(FAST FRB Key Science Project)}

\begin{abstract}
We present the scintillation velocity measurements of FRB~20201124A from the FAST observations, which reveal an annual variation.
This annual variation is further supported by changes detected in the scintillation arc as observed from the secondary spectrum. 
We attribute the annual velocity variation to the presence of a moderately anisotropic scattering screen located at a distance of 0.4$\pm$0.1~kpc from Earth. 
Our results prove that the scintillation of this FRB is mainly caused by material close to Earth on a Galactic scale.
However, scintillation observations of other FRBs may expose their surrounding environment or uncover possible orbital motion if scintillation is caused by materials in their host galaxy.
\end{abstract}


\section{Introduction} \label{sec:intro}

Fast radio bursts (FRBs) discovered by \cite{lbm+07} are a new class of astronomical source characterized by their short duration (approximately milliseconds) and highly energetic emission in the radio wavelengths (i.e., 10$^{37} - 10^{40}$~erg reported by \citealt{lwz+21}).
Observations across various frequency bands and theoretical investigations have been conducted to understand FRBs.
Notable achievements in recent years include the identification of periodic activity in FRB active window \citep{chimeper+20}, the association of FRBs with galactic magnetars \citep{brb+20, zxz+23}, the detection of short-term variations in Faraday rotation measure \citep{xnc+22}, and the identification of FRBs associated with persistent radio sources \citep{nal+22}, among others. 
Despite these advancements, FRBs' radiation mechanisms and physical origins remain enigmatic \citep{zhangb23}.

Ionized interstellar medium with varying refractive indices focuses and scatters the phase of radio waves from compact objects, e.g. pulsars and FRBs.
The interference between these scattered rays leads to modulation of the signal intensity as a function of frequency and time. 
This phenomenon is known as interstellar scintillation \citep[ISS,][]{sch68}.
ISS is typically categorized into two types: diffractive ISS \citep[DISS,][]{ric69} and refractive ISS \citep[RISS,][]{sie82, rcb84}. 
DISS results in significant modulations on short timescales, while RISS arises due to large-scale re-focusing and de-focusing of the rays, causing small modulations on long timescales \citep{ric90}.
Small-scale enhanced signals, known as scintles, can be identified in the dynamic spectra, i.e., the two-dimensional (2D) matrices of signal intensity as a function of time ($t$) and frequency ($f$).
In some cases, arcs could be identified in the secondary spectra, i.e. the 2D power spectra of dynamic spectra \citep{smc+01,wms+04, crs+06}. 
Scintillation velocity $V_{\rm{\rm{eff}}}$ related parameters, such as the scintillation timescale $\tau_{\rm{d}}$ (characteristic size of scintles in the time domain) and arc curvature $\eta$, exhibit periodic variations due to the Earth and source orbital motions \citep[i.e.,][]{lyn84, rcb+20}.
This motivates the intriguing question of whether it is possible to search for putative orbital motion of FRBs through scintillation observations.

Previous FRB scintillation studies \citep[i.e.,][]{zhz+22} focused on investigating the scintillation bandwidth $\Delta {\nu_{\rm{d}}}$. 
Measuring the scintillation timescale $\tau_{\rm{d}}$ is challenging due to the lack of a sufficient sample of bursts from active repeating FRBs that show bursts with correlated bandpasses.
\cite{nhk+22} found evidence of correlated scintillation frequency structures between two bursts from FRB~20200120E separated by 4.3 minutes, indicating the presence of scintillation timescales $\tau_{d} \geq$ 4 minutes.
\cite{mhm+22} reported the first measurement of scintillation timescale $\tau_{d}$ for FRB~20201124A using Effelsberg and the upgraded Giant Metrewave Radio Telescope. 
Moreover, the annual variation in scintillation velocity was detected based on ten scintillation velocity values, suggesting a Milky Way origin for the ISS of FRB~20201124A \citep{mbm+23}. 
Additionally, \cite{wmz+23} identified the first detection of FRB scintillation arc from FRB~20220912A, indicating the presence of a thin scattering screen in the line of sight dominating the observed scintillation phenomena.
Moreover, there is no significant difference in scintillation arc curvature of FRB~20220912A detected within the time span of 20 days.  
Readers can refer to \cite{cordes16} for more details regarding FRB scintillation.

\begin{figure*} 
    \centering
    \includegraphics[width=0.96\textwidth]{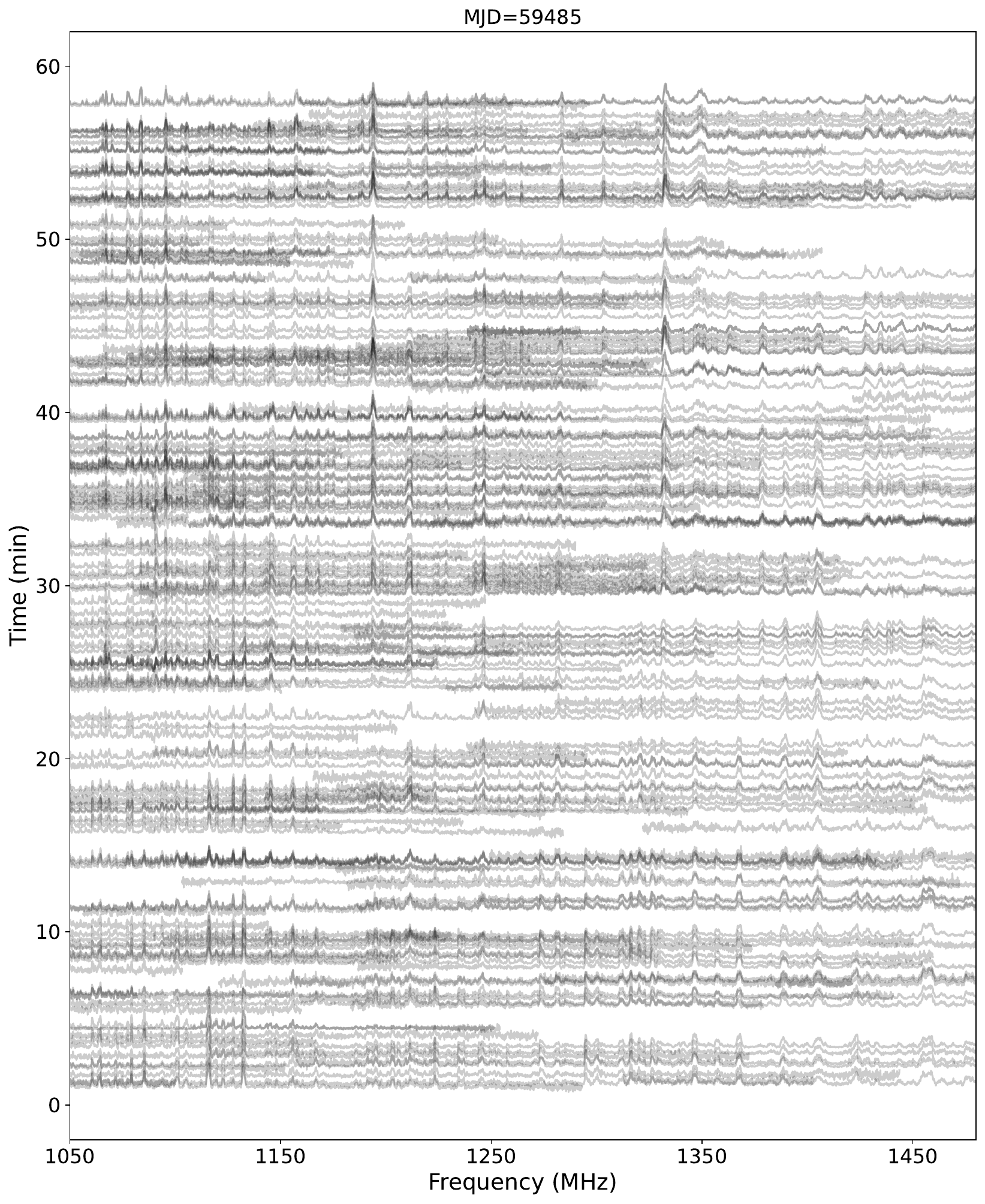}
    \caption{The dynamic spectrum of Stokes I of FRB~20201124A at MJD 59485. 
    High-frequency noises are injected for calibration during the observation's first and last minutes. 
    The same printed contrast is applied to each frequency spectrum, and the lines have transparency.}
    \label{fig:dynamicspectrum}
\end{figure*}

In this Letter, we present the annual variations in scintillation velocity and arc curvature of FRB~20201124A using the Five-hundred-meter Aperture Spherical radio Telescope (FAST). 
Section~\ref{sec:data} describes our observations and data processing procedures. 
The results are presented in Section~\ref{sec:result}. 
Finally, Section~\ref{sec:conclusion} contains the discussions and conclusions.

\section{Observations and Data analysis}
\label{sec:data}
FRB~20201124A was initially discovered by CHIME \citep{chime20201124A} and was subsequently localized to its host galaxy \citep{dbd+21}.
Subsequent FAST observations detected three active episodes, including thousands of bursts \cite{xnc+22}. 
\cite{zhz+22} studied its burst morphology.
\cite{zwf+22} studied its energy distribution.
\cite{jwx+22} did its polarimetry studies, and \cite{nzz+22} tried to find spin period and reported fine structures.

For our analysis, we utilize data obtained with FAST between April 01, 2021, and July 11, 2023 \citep{nlj+11, jyg+19, lwq+18}, in the frequency range of 1.0--1.5~GHz.
The pointing direction of FAST was R.A. = 05$^{\rm{h}}$08$^{\rm{m}}$03.51$^{\rm{s}}$, and Dec = +26$^{\circ}$03$^{'}$38.5$^{''}$, corresponding to Galactic longitude: 177.76$^{\circ}$ and Galactic latitude: -8.52$^{\circ}$.

We use search-mode filter bank data with a digital backend based on the ROACH2 board \citep{jyg+19, jth+20} without coherent dedispersion. 
The recorded data are in the 8-bit PSRFITS format \citep{hvm04} with four-channel Stokes format (XX, YY, XY, YX), the time resolution of 49.152 $\mu$s, and 4096 frequency channels corresponding to a frequency resolution of approximately 0.122~MHz.
The 19-beam receiver of FAST utilizes orthogonal linear polarization feeds. 
At the beginning of the observation, a periodic square wave with a 100\% linear polarization calibrator signal is injected, using a folded noise diode for polarization calibration. 
The cross-coupling of the two feeds is negligible ($\leq0.5$\%, \citealt{lwm+20}).
The PSR/IEEE convention \citep{vmjr10} is employed to define Stokes parameters.
We then do offline incoherent dedispersion and search for the FRB signal, using a single fixed dispersion measure (DM) value of 411 pc cm$^{-3}$ and a peak signal-to-noise ratio (S/N\footnote{The peak S/N is given by the ratio of the maximum intensity of pulse profile to the standard deviation of the off-pulse region. The time resolution of pulse profile is 49.152 $\mu$s}.) threshold of 10.0. 
It is worth noting that previous FAST FRB analyses \citep[i.e.,][]{zhz+22} employed a detection threshold of S/N = 7, enabling the detection of more bursts.
We use a higher detection threshold since studies of scintillation properties require a high signal-to-noise ratio, often from deflected ray paths with $\lesssim$1\% of the FRB's mean flux.

To mitigate radio frequency interference, we identify according to the mean and the standard deviation of the time-averaged series and then mask. 
The methodologies for generating dynamic and secondary spectra are detailed in previous work by \cite{mhm+22} and \cite{wmz+23}.
We then derive the scintillation bandwidth ($\Delta \nu_{\rm d}$) and scintillation timescale ($\tau_{\rm d}$) from the 2D autocorrelation function (ACF) of the dynamic spectrum from each observation \citep[see][for more details]{wmz+23}, but only if we have more than ten blocks of data\footnote{Each block of data is around 0.2s (4096 sampling intervals). We use the term ``block of data'' rather than burst here due to the frequent occurrence of multiple bursts within a block of data of FRB~20201124A depending on the definition of a burst.} containing brighter bursts detected (see Figure~\ref{fig:dynamicspectrum}).
This ensures the reliability of the derived scintillation parameters.
The uncertainty in the scintillation parameters is the quadrature sum of the uncertainty arising from the fitting procedure and the statistical error $\sigma_{\rm{est}}$ resulting from the limited number of scintles \citep{brg99}: 
\begin{equation}
      \rm{ \sigma_{est} = (f_{d} \boldsymbol{\times} \frac{BW_{\rm{dyn}}T_{\rm{dyn}}}{\Delta\nu_{\rm{d}} \tau_{\rm{d}}})^{-0.5}.}
          \label{eq:error}
\end{equation}
Here, $\rm{BW_{dyn}}$ and $\rm{T_{dyn}}$ are the observing bandwidth and length, respectively, and $f_{d}$ (= 0.4) is the filling factor.
$\sigma_{\rm{est}}$ is 2--10\% mainly depending on the observing length.

\section{Results}
\label{sec:result}
We analyze all the datasets of FRB~20201124A obtained from FAST and identify a total of 51 observations with more than ten selected blocks of data.
In Table~\ref{tab:summaries}, we provide the corresponding Modified Julian Dates (MJDs), observing lengths, the number of blocks of data, scintillation bandwidth $\Delta \nu_{\rm{d}}$, and scintillation timescale $\tau_{\rm{d}}$ of the selected observations.

\begin{figure}
    \centering
    \includegraphics[width=0.49\linewidth]{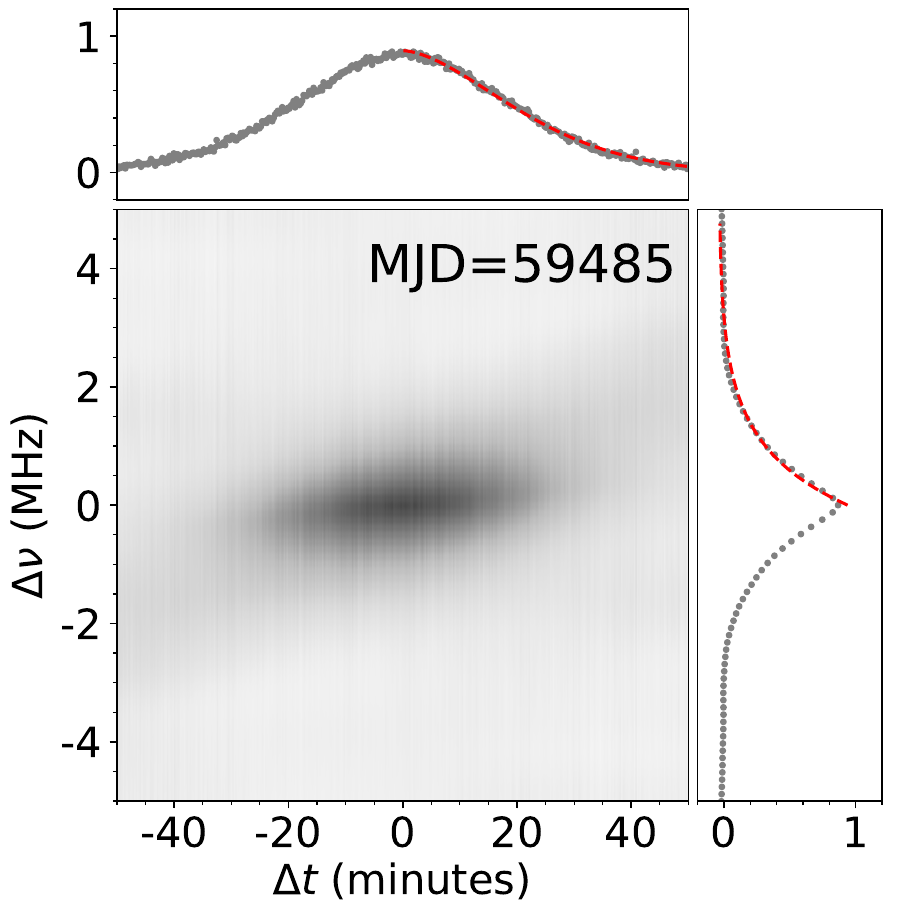}
    \includegraphics[width=0.49\linewidth]{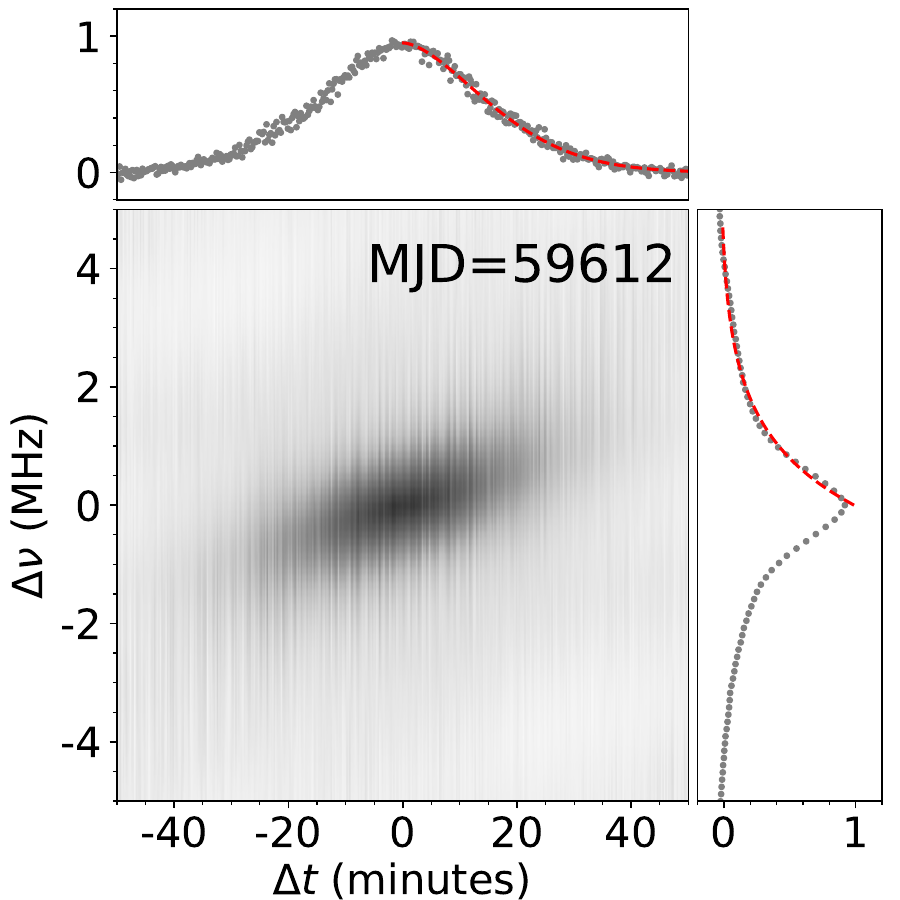}  \\
    \includegraphics[width=0.49\linewidth]{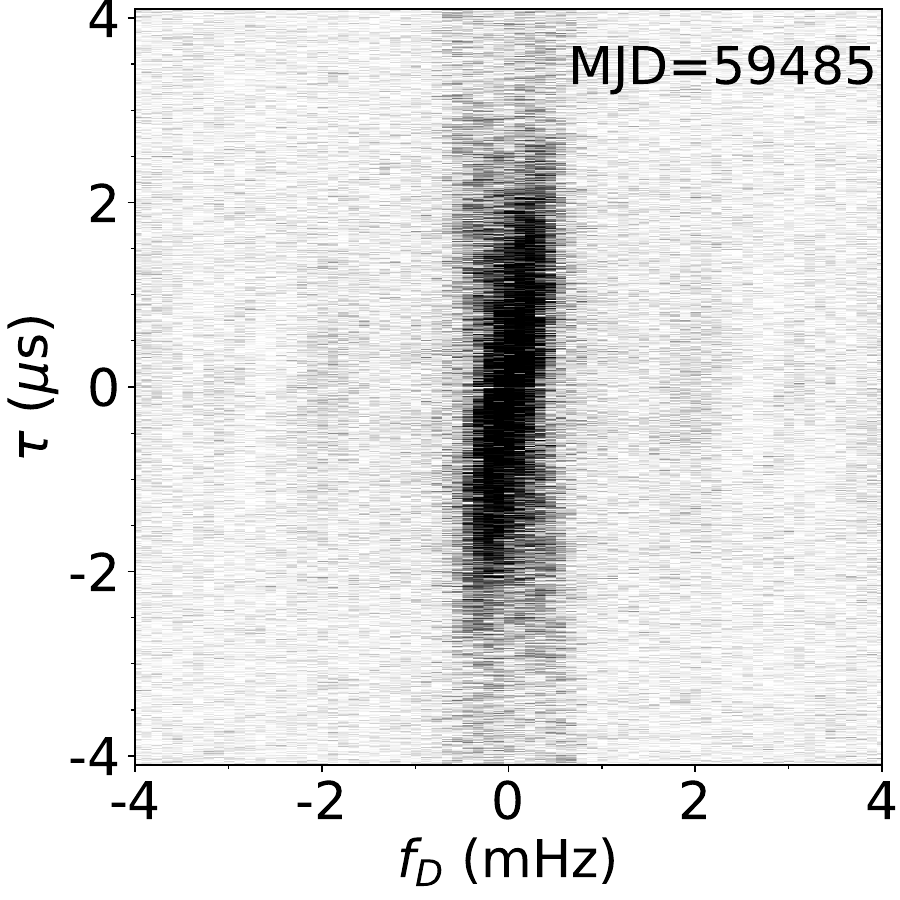}
    \includegraphics[width=0.49\linewidth]{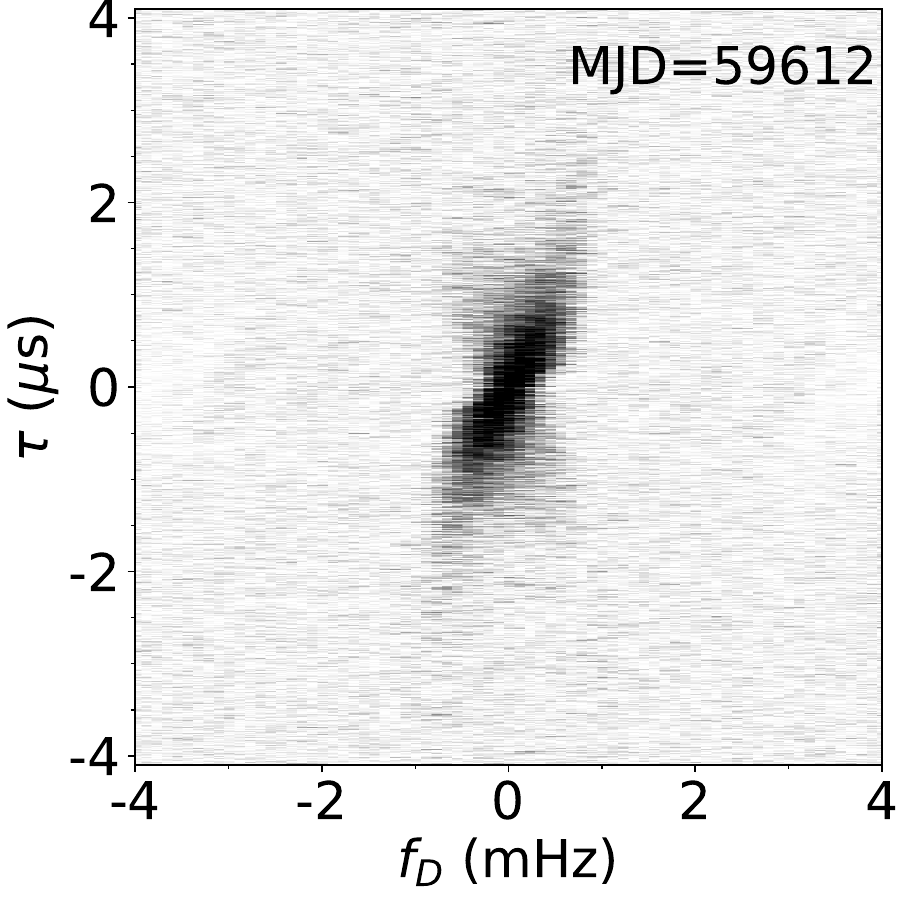} 
    \caption{(Upper panels) The 2D ACFs of FRB~20201124A with FAST on MJDs 59485 and 59612. In the two smaller side plots, the gray points are the 1D ACFs at zero frequency and time lag, the red dashed curves are the best fits from which the scintillation timescale $\tau_{\rm{d}}$ and scintillation bandwidth $\Delta \nu_{\rm{d}}$ are derived, respectively. (Bottom Panels) The corresponding secondary spectra.}
    \label{fig:acf_ss}
\end{figure}

We present the 2D ACFs and the secondary spectra of MJD 59485 and 59612 in Figure~\ref{fig:acf_ss} as examples.
Both of these examples exhibit tilts caused by refraction in their 2D ACFs.
The presence of asymmetrical scintillation arcs provides further evidence for the existence of a larger-scale gradient in phase \citep{crs+06}.
\begin{table*}
\begin{center}
\begin{threeparttable}
\caption{Summary of 51 FAST observations and scintillation properties of FRB~20201124A}
\footnotesize 
\begin{tabular}{lccccc}
\hline \hline
MJD & Length (hrs) & $N_{\mathrm{blocks}}$ & $\tau_{\rm{d}}$ (mins) & $\Delta \nu_{\rm{d}}$ (MHz) & Arc Detection \\
\hline
59314 & 2.0 & 59 & 17.7$\pm$0.6 & 0.9$\pm$0.1 &  \\
59315 & 2.0 & 59 & 17.1$\pm$0.7 & 1.2$\pm$0.1 &   \\
59316 & 2.0 & 44 & 15.9$\pm$0.6 & 0.9$\pm$0.1 &  \\
59319 & 1.0 & 19 & 15.8$\pm$0.8 & 0.5$\pm$0.1 &   \\
59320 & 1.9 & 25 & 16.4$\pm$0.7 & 0.5$\pm$0.1 &   \\
59321 & 1.5 & 19 & 15.7$\pm$1.0 & 0.7$\pm$0.1 &   \\
59323 & 2.0 & 46 & 18.4$\pm$0.8 & 0.7$\pm$0.1 &   \\
59324 & 1.7 & 24 & 12.1$\pm$0.9 & 1.4$\pm$0.1 &   \\
59325 & 0.5$\times$4 & 29 & 13.5$\pm$0.8 & 0.9$\pm$0.1 &   \\
59326 & 0.5$\times$3 + 0.8$\times$1 & 44 & 12.7$\pm$0.8 & 1.2$\pm$0.1 &   \\
59327 & 0.5$\times$4 & 24 & 15.6$\pm$0.9 & 0.5$\pm$0.1 &   \\
59328 & 0.5$\times$4 & 34 & 12.0$\pm$0.7 & 0.5$\pm$0.1 &   \\
59329 & 1.5 & 21 & 14.1$\pm$0.9 & 0.8$\pm$0.1 &   \\
59330 & 0.5$\times$4 & 29 & 14.5$\pm$0.9 & 0.7$\pm$0.1 &   \\
59331 & 0.5$\times$4 & 28 & 10.7$\pm$0.8 & 0.5$\pm$0.1 &   \\
59334 & 0.5$\times$4 & 43 & 12.8$\pm$0.8 & 0.7$\pm$0.1 &   \\ 
59336 & 0.5$\times$4 & 19 & 13.4$\pm$1.4 & 0.5$\pm$0.1 &   \\
59337 & 0.5$\times$4 & 37 & 15.2$\pm$1.4 & 0.6$\pm$0.1 &   \\
59347 & 0.5$\times$4 & 13 & 13.4$\pm$1.6 & 0.7$\pm$0.1 &   \\
59348 & 0.5$\times$4 & 21 & 10.7$\pm$1.1 & 1.1$\pm$0.1 &  \\
59349 & 0.5$\times$4 & 15 & 9.5$\pm$1.3 & 0.5$\pm$0.1 &   \\
59350 & 0.5$\times$4 & 20 &10.4$\pm$0.7 & 1.4$\pm$0.1 &   \\
59351 & 0.5$\times$4 & 23 & 11.1$\pm$1.2 & 0.6$\pm$0.1 &   \\
59352 & 0.5$\times$4 & 26 & 9.5$\pm$1.0 &  0.7$\pm$0.1 &   \\
59482 & 1.0 & 15 & 30.1$\pm$2.2 & 0.4$\pm$0.1 &   \\
59483 & 1.0 & 40 & 34.5$\pm$1.8 & 0.6$\pm$0.1 &   \\
59484 & 1.0 &127 & 28.0$\pm$1.0 & 0.5$\pm$0.1 & Yes   \\
59485 & 1.0 &283 & 25.8$\pm$1.0 & 0.7$\pm$0.1 & Yes  \\
59612 & 2.0 &187 & 20.1$\pm$0.8 & 0.7$\pm$0.1 & Yes  \\
59613 & 2.0 & 150& 17.8$\pm$0.4 & 0.6$\pm$0.1 & Yes  \\
59614 & 2.0 & 61 & 16.4$\pm$0.6 & 0.9$\pm$0.1 & Yes  \\
59615 & 1.0 & 54 & 17.8$\pm$0.6 & 0.7$\pm$0.1 & Yes  \\
59616 & 1.0 & 37 & 19.1$\pm$0.8 & 0.8$\pm$0.1 &  \\
59619 & 1.0 & 34 & 24.9$\pm$1.5 & 0.8$\pm$0.1 &  \\
59621 & 1.0 & 30 & 21.9$\pm$1.5 & 1.3$\pm$0.1 &    \\
59622 & 1.0 & 27 & 27.9$\pm$2.2 & 2.1$\pm$0.2 &   \\
59624 & 1.0 & 18 & 24.9$\pm$1.5 & 0.4$\pm$0.1 &   \\
59625 & 1.0 & 24 & 19.5$\pm$1.5 & 0.7$\pm$0.1 &   \\
59628 & 0.33& 16 & 21.1$\pm$7.9 & 0.7$\pm$0.1 &   \\
59629 & 0.33& 15 & 22.9$\pm$2.6 & 1.5$\pm$0.2 &   \\
59630 & 0.33& 29 & 22.0$\pm$1.7 & 0.7$\pm$0.1 &   \\
59631 & 0.33& 32 & 21.6$\pm$2.3 & 1.4$\pm$0.1 &   \\
59632 & 1.0 & 66 & 21.7$\pm$1.4 & 0.7$\pm$0.1 &   \\
59633 & 0.33& 29 & 18.5$\pm$1.3 & 0.9$\pm$0.1 &   \\
59634 & 1.0 & 72 & 23.9$\pm$1.5 & 0.6$\pm$0.1 &   \\
59636 & 1.0 & 46 & 24.2$\pm$1.1 & 0.6$\pm$0.1 &   \\
59637 & 0.33& 14 & 23.2$\pm$2.6 & 0.6$\pm$0.1 &   \\
59638 & 0.33& 12 & 22.3$\pm$2.9 & 1.2$\pm$0.1 &   \\
59642 & 0.33& 12 & 17.6$\pm$4.0 & 0.6$\pm$0.1 &   \\
59657 & 0.33& 11 & 24.6$\pm$2.6 & 1.3$\pm$0.1 &   \\
59658 & 0.33& 15 & 39.0$\pm$7.7 & 0.9$\pm$0.1 &  \\
\hline
    \end{tabular}
    \label{tab:summaries}
\begin{tablenotes}
\small
\item {\bf Notes:} Given are the Modified Julian Day (MJD), the observation length, $N_{\mathrm{blocks}}$: the number of blocks of data (the integration time of each blocks of data is $\sim$0.2s, and some of blocks of data have multiple bright bursts.), the derived scintillation timescale $\tau_{\rm{d}}$ and bandwidth $\Delta \nu_{\rm{d}}$, and the note whether scintillation arc is detected.
\end{tablenotes}
\end{threeparttable}
\end{center}
\end{table*}

\begin{figure}
    \centering
    \includegraphics[width=0.99\linewidth]{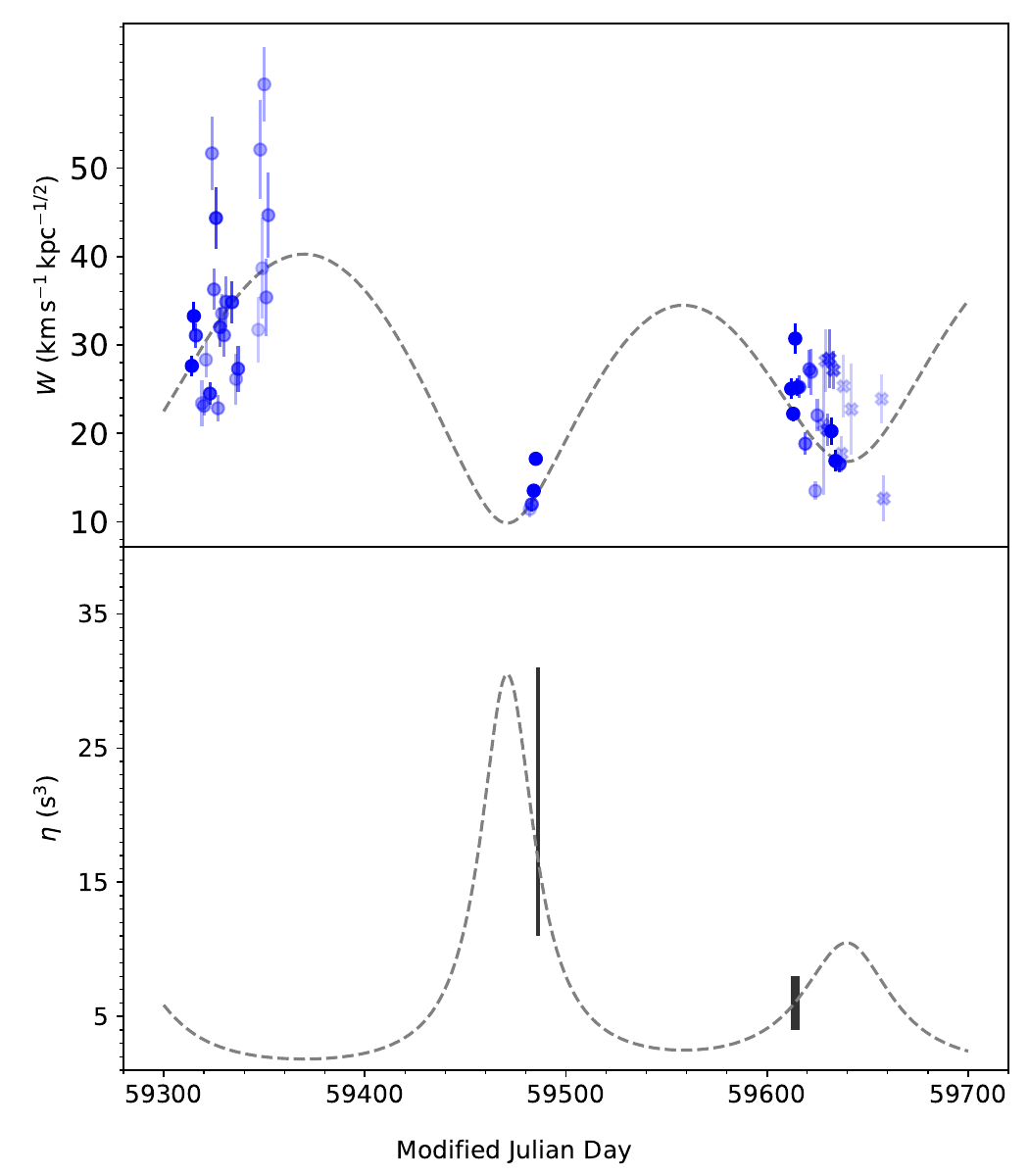}
    \caption{(Upper panel) Measurements of the re-scaled scintillation velocity $W$ of FRB~20201124A. The dashed curve represents the best-fitting anisotropic scattering model described in section~\ref{sec:model} using MCMC. The brightness of $W$ depends on the number of blocks of data of each observation. (Lower panel) The predicted scintillation arc using the best-fitting anisotropic scattering model. The black-filled areas represent the scintillation arc curvature range.}
    \label{fig:mcmc}
\end{figure}

\subsection{Scintillation Velocity}
The predominant scattering model employed is the thin screen model \citep{wil72, wil73}, which postulates that scattering occurs within a narrow screen oriented along the direction of propagation.
The validity of the thin screen model was corroborated by the detection of very sharp scintillation arcs \citep[i.e.,][]{smc+01, crs+06, pl14}.
In our observations, the secondary spectra of FRB~20201124A shown in Figure~\ref{fig:acf_ss} are of the fuzzy type and are more suggestive of a non-thin screen geometry. For simplicity, the thin screen model is adopted.

In the study of \textbf{time} variations in scintillation phenomena, a crucial parameter is the effective transverse line-of-sight velocity $V_{\rm eff}(s)$ across the scattering medium at fraction position $s$ that is measured from the FRB position at $s$=0 to $s$=1 at the observer's position. 
This velocity is a linear combination of the velocities of the FRB, Earth, and the ionized interstellar medium (IISM) in the form of \citep{cr98}:
\begin{equation}
\label{eqn:veff1}
V_{\rm eff}(s) = (1 - s)(V_{\rm {FRB,T}} + V_{\rm FRB,O}) + sV_{\rm E} - V_{\rm IISM}(s)
\end{equation}
where $V_{\rm {FRB,T}}$, $V_{\rm {FRB,O}}$, $V_{\rm E}$ and $V_{\rm IISM}$ are the velocities of the FRB proper motion, the possible orbital motion of the FRB, earth motion and the IISM motion, respectively.
Each velocity component comprises two parts: $v_{\mathrm{\alpha}}$ along R.A. and $v_{\mathrm{\delta}}$ along Dec.
In the case of the Milky Way origin of FRB, the fractional distance s$\sim$1.
Consequently, the contributions from the FRB velocities can be considered negligible. 

In this Letter, we adopt a parameter $W$ instead of effective scintillation velocity $V_{\rm{eff}}$.
This choice allows us to isolate the unknown variables from the directly measurable parameters \citep{mbm+23}:
\begin{equation}
\label{equ:w}
\begin{aligned}
W_{\mathrm{km/s/\sqrt{kpc}}} \equiv \frac{V_{\mathrm{ISS}}}{\sqrt{D_{\rm{s}}}} 
  &= \frac{V_{\mathrm{eff}}}{s\sqrt{D_{\rm{s}}}}
  \approx \frac{V_{\mathrm{eff}}}{\sqrt{D_{\rm{s}}}}, (\rm{theory}) \\
  &\approx 27800 \frac{\sqrt{2\Delta \nu_{\rm{d}}}}{f\tau_{\rm{d}}}, (\rm{observation})
\end{aligned}
\end{equation}
where V$_{\mathrm{ISS}}$ is the scintillation velocity in km/s \citep{cr98}, D$_{\rm{s}}$ is the distance of scattering screen from Earth in the unit of kpc, the scintillation bandwidth $\Delta \nu_{\rm{d}}$ is expressed in the unit of MHz, the observing frequency $f$ is in the unit of GHz, and the scintillation timescale $\tau_{\rm{d}}$ is represented in the unit of second. 

\subsection{Anisotropic scattering screen model}
\label{sec:model}
Observational phenomena such as extremely scattering events \citep{fdjh87, cks+15}, the scintillation behavior of B0834+06 \citep{bmg+10}, and flatter power law index at lower frequencies \citep{wcv+23} strongly suggest the presence of inhomogeneous and anisotropic scattering screens.
We consider the scattered image of a point source produced by the screen that is an ellipse with the axial ratio A$_{\rm{r}}$ and $\phi_{\rm{A}_{\rm{r}}}$ defined clockwise from R.A. \citep{cmr+05}.
\cite{rcn+14} further introduced a parameterization of the quadratic coefficients in terms of $R=(A_r^2 - 1)/(A_r^2 + 1)$, which ranges between 0 (isotropic screen) and 1 (1D screen).
The coefficients are
\begin{equation}
\label{eqn:anisotropy}
\begin{aligned}
a &=  \left[1 - R\cos{\left(2\phi_{\rm A_r}\right)}\right]/\sqrt{1 - R^2}, \\ 
b &=  \left[1 + R\cos{\left(2\phi_{\rm A_r}\right)}\right]/\sqrt{1 - R^2}, \\ 
c &=  - 2R\sin{\left(2\phi_{\rm A_r}\right)}/\sqrt{1 - R^2}.
\end{aligned}
\end{equation}
Eventually, the theoretical model of the effective scintillation velocity is then given by \citep{rcn+14}
\begin{equation}
\label{eqn:veff_anisotropic}
V_{\rm eff}(s) = \sqrt{a v_{\mathrm{\alpha}}(s)^2 + b v_{\mathrm{\delta}}(s)^2 + c v_{\mathrm{\alpha}}(s)v_{\mathrm{\delta}}(s)}.
\end{equation}

\subsection{Interpreting Scintillation Variation}
We compute the parameter  $W$ with Equation~\ref{equ:w} using the mean value of scintillation bandwidth $\Delta \nu_{\rm{d}}$.
We employ this approach for two reasons.
Firstly, due to the presence of phase gradient variation manifested in tilted ACFs and the dynamic asymmetry of scintillation arcs, as depicted in Figure~\ref{fig:acf_ss}.
Secondly, individual bursts do not cover the full frequency band.
These factors can lead to complications in scintillation parameter measurements. 
Considering that the scintillation bandwidth $\Delta \nu_{\rm{d}}$ strongly depends on frequency, we tend to use its mean value. 
The resulting time series of $W$ is illustrated in Figure~\ref{fig:mcmc}. 
It is worth noting that the scintillation arc curvature also contains information about $W$.
However, precise arc curvature measurements are hindered by the presence of diffuse arcs with larger curvature. 
Hence, in this letter, we derive $W$ only from the perspective of $\Delta \nu_{\rm{d}}$ and $\tau_{\rm{d}}$.

\begin{figure}
    \centering
    \includegraphics[width=0.99\linewidth]{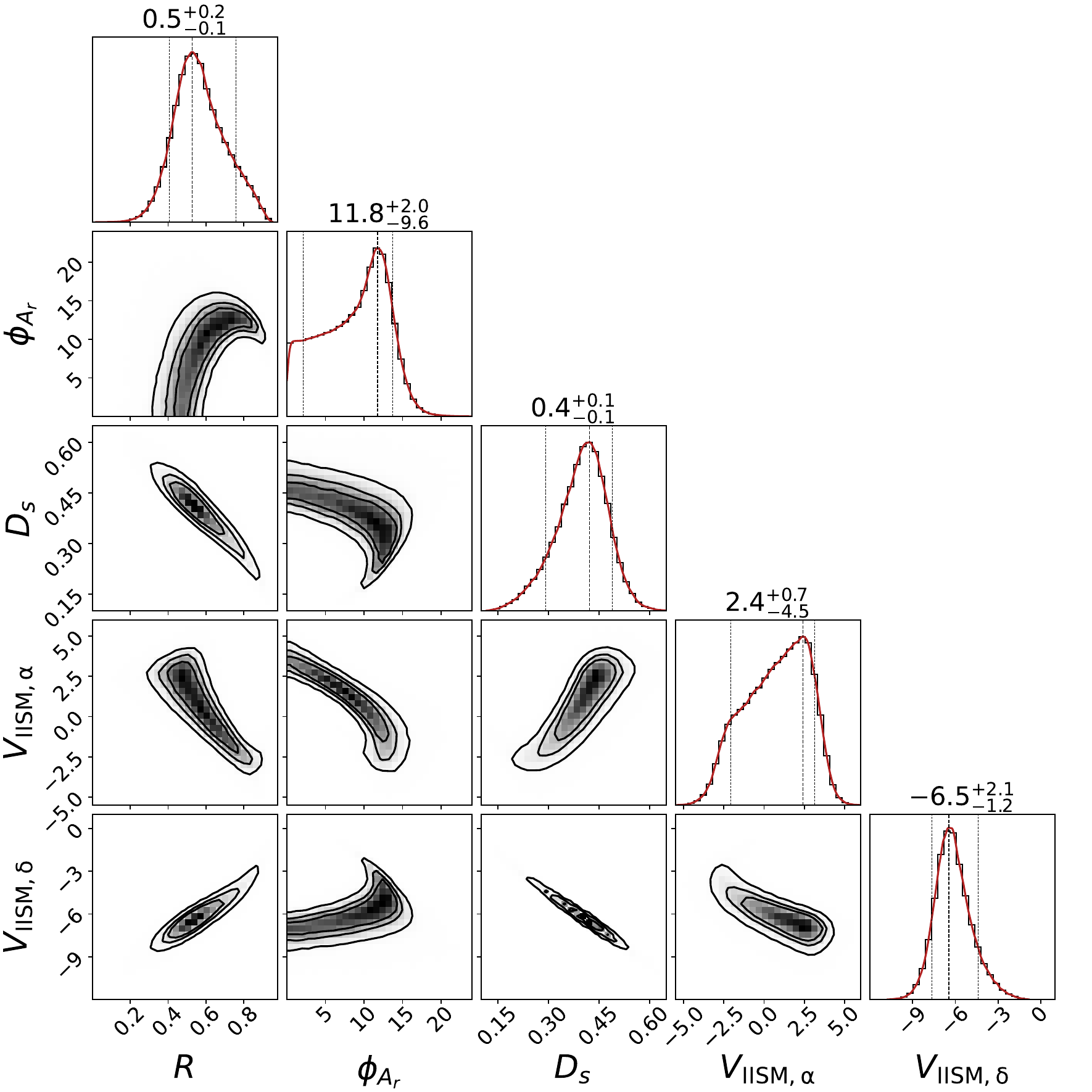}
    \caption{The posterior probability distribution of all fitted parameters. In the 1-D histograms, the red curves represent the kernel-density estimate smoothed versions of the distributions. From left to right, three black dashed lines denote the 10\% fractional percentiles, the most likely values, and the 90\% fractional percentiles, respectively. The most likely values and the upper/lower errors are indicated at the top of the 1-D histograms.}
    \label{fig:corner}
\end{figure}

We model the time series of the parameter $W$ using Markov chain Monte Carlo (MCMC) with EMCEE \citep{fhlg13}.
All fitting parameters have uniform priors, and their posterior distributions are presented in Figure~\ref{fig:corner} using corner.py \citep{corner}. 
The best-fit parameters for $R$, $\phi_{\rm{A}_{\rm{r}}}$, $D_{\rm{s}}$, $V_{\rm{IISM, \alpha}}$, and $V_{\rm{IISM, \delta}}$ are 0.5$^{+0.2}_{-0.1}$, 11.8$^{+2.0}_{-9.6}$~degrees, 0.4$\pm{0.1}$~kpc, 2.4$^{+0.7}_{-4.5}$~km/s, -6.5$^{+2.1}_{-1.2}$~km/s, respectively.
The results of our model fitting are depicted in Figure~\ref{fig:mcmc}.
The axial ratio A$_{r}$ of the spatial diffraction pattern is 1.7$^{+0.7}_{-0.2}$, indicating a moderately anisotropic scattering screen responsible for FRB~20201124A ISS. 
Additionally, our result of $D_{\rm{s}}$ = 0.4$\pm$0.1 kpc is consistent with the results of \cite{mbm+23}.

Using the best-fitted anisotropic scattering screen model, we also predict the scintillation arc curvature $\eta = \lambda^{2}/2\rm{cW}^{2}$ (see Figure~\ref{fig:mcmc}), where $\lambda$, $\rm{c}$ are the observing wavelength and the speed of light, respectively \citep[i.e.,][]{mma+22}. 
We also show the range of scintillation arc curvature.
The $\eta$ measurements at MJDs 59484 and 59485 are larger than those during MJDs~59612 to 59615.
This variation in scintillation arc curvature also supports the Milky Way origin of FRB~20201124A ISS. 
In principle, the presence of a scintillation arc should be more apparent during the first active episode (MJD~59314 - 59352) due to the smaller predicted scintillation arc curvature, $\eta$. 
The absence of such an arc during these epochs can be attributed to the limited number of blocks of data, in other words, by the limited event rate of bright bursts.

\section{Discussions and Conclusions}
\label{sec:conclusion}
Using the FAST data set for FRB~20201124A, our results show sustained frequency structure ($\sim$1 MHz, see Figure~\ref{fig:dynamicspectrum}) across multiple bursts due to diffractive interstellar scintillation and demonstration of a scintillation arc.

There are common features in the frequency spectra from the bursts near each other, as shown in Figure~\ref{fig:dynamicspectrum}.
However, it is not a direct signal that there is no variation in the intrinsic emission of FRB involved.
Additional frequency structure caused by the intrinsic emission from pulsars, a typical coherent source, was first detected and analyzed by \cite{cbh+04}. 
The radiation mechanism for FRB emission should be coherent, given the effective brightness temperature needed \citep[i.e.,][]{zhangb23}.
\cite{nhk+22} showed that the correlation coefficient of Stokes I depends on the degree of polarization of FRB.
A more quantitative analysis of the correlation of frequency spectra and the degree of polarization is deferred to a future paper.

We model the annual variation in scintillation velocity of FRB~20201124A with an anisotropic scattering screen model based on the largest sample of FRB scintillation measurements with FAST, up to date. 
Our findings strongly indicate that the observed ISS phenomena of FRB~20201124A are predominantly influenced by the IISM with a transverse velocity of approximately 7~km/s located at 0.4$\pm$0.1~kpc away from Earth.
This is further supported by the variation in scintillation arcs detected within the time intervals of MJDs 59484-59485 and 59612-59615.
The annual orbital motion of Earth is responsible for the yearly variation in the scintillation velocity.

This phenomenon of annual variation in scintillation has been observed in eight quasars: 0917+642 \citep{rwk+01}, S4 0954+65 \citep{mkl+12}, 1156+295 \citep{llm+13}, B1257$-$326 \citep{bjl+03}, B1322-110 \citep{brs+19}, B1519-273 \citep{lrr08}, J1726+0639 \citep{bts+22}, J1819+3845 \citep{dd03}, nine pulsars: J0437$-$4715 \citep{rcb+20}, J0613$-$0200 \citep{msa+20}, J0636+5128 \citep{lmv+23}, J0835$-$4510 \citep{xsl+23}, J1136+1551 \citep{mzs+22}, J1141$-$6545 \citep{rch+19}, J1509+5531 \citep{smw+22}, J1643$-$1224 \citep{mma+22}, and J1909$-$3744 \citep{ars23}.
The annual variation in ISS of FRB~20201124A demonstrates FRB's potential for exploring the material's properties in the vicinity of Earth. 
By utilizing multi-epoch data, the detection of annual scintillation variations from compact sources can provide valuable insights into the local IISM for many lines of sight and further enable nearby IISM reconstruction in the future \citep[i.e.,][]{occ+24}. 

FAST, as the most sensitive single-dish telescope in the world, has the ability to detect a sufficient number of bright bursts to obtain the scintillation timescale $\tau_{\rm{d}}$, even to resolve the scintillation arc.
The annual variation in the scintillation arcs yields robust constraints on the properties of the scattering screen if the arcs are well-defined.
However, in the case of FRBs, a high event rate of bright bursts is crucial to search for the scintillation arc effectively. 
A similar scintillation analysis can be conducted for FRBs exhibiting exceptional activity within the sky coverage of FAST, such as FRB~20121102 \citep{lwz+21}. 
The annual variation of scintillation would improve our understanding of Milky Way IISM distributions at high latitudes.
A non-annual periodic variation in scintillation velocity would be even more remarkable, as it could signify the orbital motion of the progenitor of the FRB.

\section*{Data Availability}
Raw data are published by the FAST data center and can be accessed there.

\section*{acknowledgments}
We thank the anonymous referee for the constructive comments and suggestions, which helped us to improve this paper.
This work made use of data from the FAST. 
FAST is a Chinese national mega-science facility built and operated by the National Astronomical Observatories, Chinese Academy of Sciences. 
Ziwei Wu is supported by the Project funded by China Postdoctoral Science Foundation No. 2023M743517.
This work is supported by the National SKA Program of China No. 2020SKA0120200, 2020SKA0120100, 
CAS Project for Young Scientists in Basic Research YSBR-063,
the National Nature Science Foundation grant No. 12041303, 11988101, 11833009, 11873067, 12041304, 12273009, 12203045, 12203013, 12303042, and  
the National Key R$\&$D Program of China No. 2017YFA0402600, 2021YFA0718500, 2017YFA0402602, 2022YFC2205203, 
the CAS-MPG LEGACY project, the Max-Planck Partner Group, the Key Research Project of Zhejiang Lab no. 2021PE0AC0 and also the Western Light Youth Project of Chinese Academy of Sciences. 

\newpage
\bibliography{journals,modrefs,psrrefs,crossrefs}
\bibliographystyle{aasjournal}

\end{document}